# Optical Tweezers as a Micromechanical Tool for Studying Defects in 2D Colloidal Crystals (Invited Paper)

Sungcheol Kim, Lichao Yu, Stephanie Huang, Alexandros Pertsinidis*, and

Xinsheng Sean Ling

*Department of Physics, Brown University, Providence, RI 02912*

**ABSTRACT**

This paper reports on some new results from the analyses of the video microscopy data obtained in a prior experiment on two-dimensional (2D) colloidal crystals. It was reported previously that optical tweezers can be used to create mono- and di-vacancies in a 2D colloidal crystal. Here we report the results on the creation of a vacancy-interstitial pair, as well as tri-vacancies. It is found the vacancy-interstitial pair can be long-lived, but they do annihilate each other. The behavior of tri-vacancies is most intriguing, as it fluctuates between a configuration of bound pairs of dislocations and that of a locally amorphous state. The relevance of this observation to the issue of the nature of 2D melting is discussed.

**INTRODUCTION**

A crystal differs from its melted liquid state in a fundamental way – it possesses long-wavelength shear rigidity and broken symmetry, the latter manifests in the striking appearance of long-range order [1]. How a crystal loses its long-range order and shear rigidity has been a subject of longstanding interest in physics. Many issues are unresolved with regard to the physics of structural defects such as vacancies, interstitials, dislocations, etc., and their roles in crystal melting.

2D colloidal crystals provide an excellent model system for studying the physics of defects. In this system, the real-time trajectories of individual colloidal particles can be tracked by using an optical microscope, recorded and analyzed using modern digital video technology. As a result, one can carry out detailed studies of the dynamics of defects in this system, as was demonstrated in the previous reports [2,3,4] on mono- and di-vacancies. Here, we report the results from analyses of the remaining video data of the same experiment.

**EXPERIMENTAL METHOD**

The details of the experiment were given in [4]. Here we recall some of the key details. The colloid sample was from a commercial source (Duke Scientific, sample No. 5036, polydispersity 1%), it contains 1%vol. aqueous suspension of 0.360 $\mu$m (in diameter) negatively charged polystyrene-sulfate microspheres. It was completely deionized by ion exchange, resulting in an ionic conductivity of $\sigma$ = 2.5 $\mu$S cm$^{-1}$. We estimated that the effective particle charge of $Z^*$ ≈ 1650 and the Debye-Huckel screening length of $\kappa^{-1}$ ≈ 390 nm. The experiment was conducted at room temperature T = 300 K.

To create a 2D colloidal crystal, a monolayer of microspheres was confined between two fused silica substrates with spacing ≈ 2 $\mu$m, resulting in a 2D crystal with a lattice constant $a$ ~ 1.1$\mu$m. The 2D crystal has an elastic energy per unit cell ~ 350 $k_BT$ (estimated by measuring the mean-square-displacement of a particle on the lattice).

The optical tweezers setup was also described in detail previously [4]. It was constructed using rotatable mirrors such that the optical trap can be moved around across the entirely field of view (~ 50 $\mu$m x 50 $\mu$m). The laser has a wavelength of 514 nm, and output power of up to 1 W.

The optical system is a Zeiss Axiovert 135 inverted microscope with a 100x oil immersion objective (N.A.=1.3). The imaging was done in the standard bright-field mode. The optical trap was created by feeding the expanded laser beam from the back aperture of the objective.

The images were recorded using a CCD camera (Sony SSC-M370) and stored on video tapes using a Sony SVO-9500MD recorder. The image analyses were done using a National Instrument frame grabber and the IDL image analysis software.

**PROCESSES OF VACANCY FORMATION**

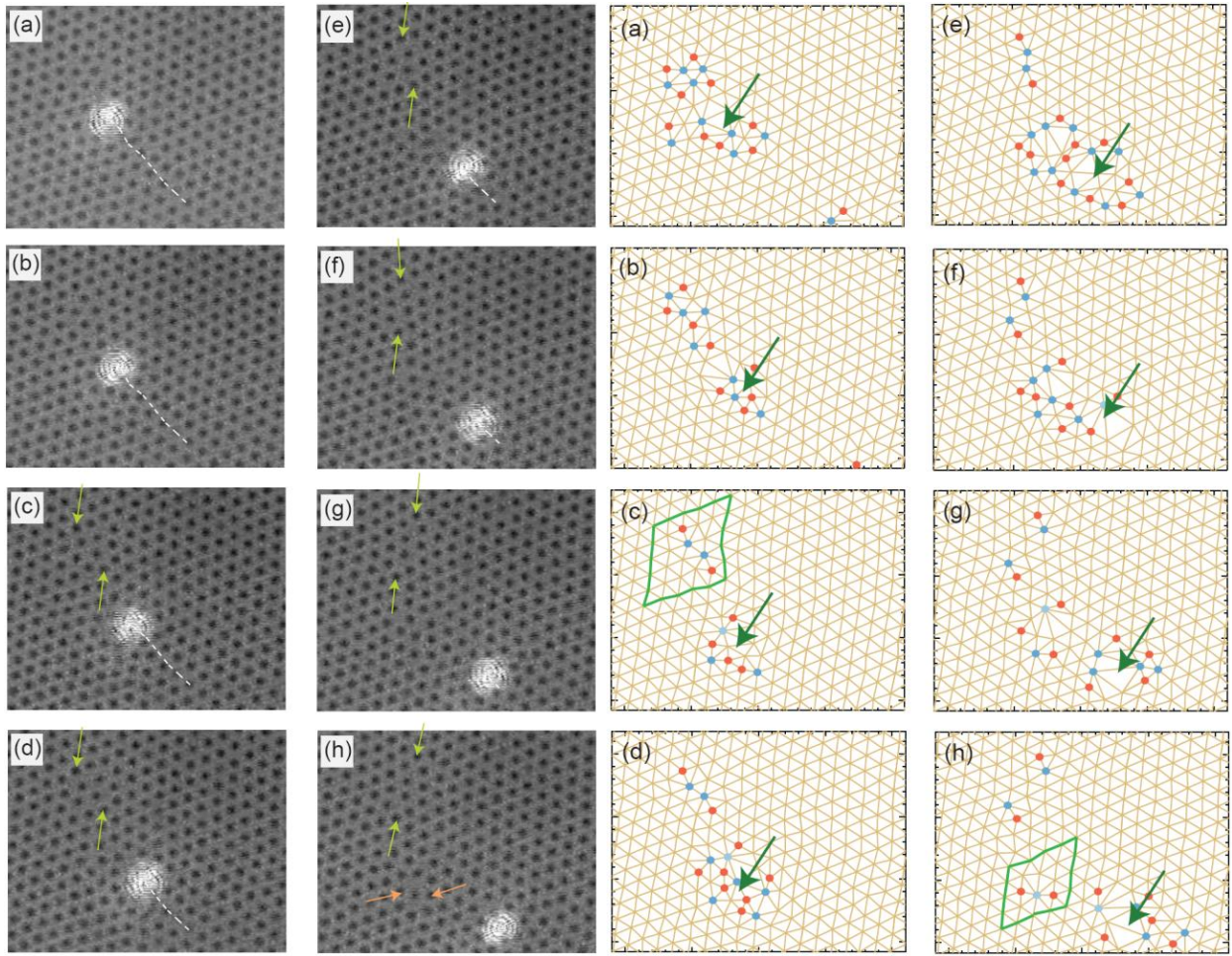

**Figure 1**: (Color online) (Left) Time-lapsed video images of optical tweezers dragging a particle in a 2D colloidal crystal. (Right) Delaunay triangulation diagrams of the corresponding video images on the left. The thick green arrows in the right represent the locations of the optical tweezers in each frame. Red and blue dots represent 5- and 7-fold coordinated particles, and are thus disclinations. Topologically, a pair of disclinations make a dislocation, and a pair of dislocations give a vacancy, as marked in (c) and (h) in the green boxes.

For this experiment, the 2D region in the sample cell extends over several hundred lattice constants. The optical tweezers experiments were conducted on a perfectly crystalline region of size 40 × 40 lattice constants. This sample region was achieved by removing the impurities using the optical tweezers. In this experiment, a laser power ≈50 mW was used to form the optical trap. The trap is strong enough to allow us to trap and drag a particle to move in any direction in the colloidal lattice. Figure 1(left) shows a few selected video images of the experiment, in sequential order, (a-h). The white dotted lines indicate the traces of the optical tweezers. The green arrows mark 5-fold coordinated particles (disclinations) which are also the cores of a pair of edge dislocations. In the previous reports [2,3,4], we showed that optical tweezers can be a convenient tool for creating point defects in a 2D colloidal crystal. It was found that the vacancies diffuse along preferential directions along the crystal axes. For di-vacancies, excitations into clearly identifiable dislocation pairs were observed, this effect allowed Peierls barrier for dislocation motion to be observed directly for the first time [2]. Here we show additional findings in this experiment.

In Figure 1, from (a) to (c), one can see that the effect of the optical tweezers dragging a particle is locally creating a disordered region, similar to shear-induced melting, which then undergoes a re-ordering, leaving only a pair of well defined dislocations, as in (d). A simple counting of particles in the green box in Fig. 1 (right, c) will show that this is a di-vacancy. Along the moving track of the optical tweezers, another vacancy is formed. As shown by the red arrows in Fig. 1 (left, h), and confirmed by counting the particles in the green box in Fig.1 (right, h), a mono-vacancy is formed. The angle between the lattice axis and the direction of optical tweezers is ~ 15 degrees in (a),(b),(c),(d) and changes to ~ 5 degrees in (e),(f),(g),(h). In this event, a single movement of the optical tweezers creates two vacancies. We found that one can create a tri-vacancy (see below) by running the optical tweezer ~10 degrees off the lattice axis.

**CREATION OF VACANCY-INTERSTITIAL PAIR**

Occasionally, the trapped particle in the optical tweezers pops out of the trap, leaving an interstitial in the lattice. Figure 2 shows an example of a vacancy-interstitial pair created with such a simple method.

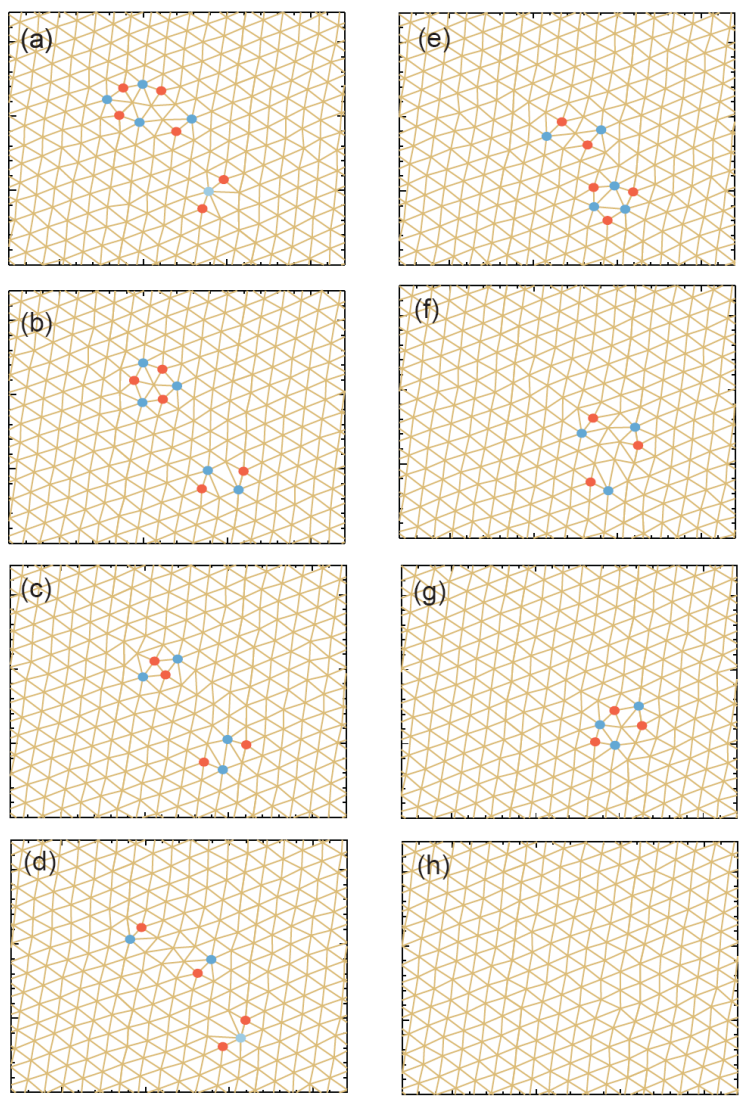

**Figure 2**: (Color online) Time-lapsed images (in Delaunay triangulation) of a vacancy-interstitial pair.

Immediately after the laser trap moved out of the field of view, in the upper left corner of Fig. 2 (a), an interstitial causes several nearby particles to have mis-coordinations, i.e. disclinations. However, the defect quickly settles into a few well defined configurations. The interstitial continuously changes its configurations, similar to the behavior of vacancy [2]. In frame Fig. 2 (b) the interstitial is in three-fold symmetry and it then changes to two-fold symmetry in frame Fig. 2 (c). In frame Fig. 2 (d), the interstitial clearly splits into a pair of dislocations separated by several lattice constants. The energy differences between these configurations are of the same order of the thermal energy $k_BT$.

It is easily observed that the interstitial diffuses faster than a nearby mono-vacancy, in lower-right of the image frames, similar to the observation of numerical studies [5]. It is apparent that there is weak attraction between the diffusing vacancy and interstitial. At frame 2(h) the vacancy and interstitial annihilate each other, leaving a perfect lattice.

To quantify the possible interactions between the mono-vacancy and the interstitial in Fig. 2, we measured the distance between the defects. The position of each defect is defined as the center-of-mass position of the disclinations associated with each defect. From the positions of each defect, we can determine the relative distance between a pair of defects as a function of time. This is done for the mono-vacancy-di-vacancy pair in Figure 1 as well, and for the vacancy-interstitial pair in Figure 2. For comparison, the results of these calculations are shown in Figure 3 below.

**EFFECTIVE INTERACTIONS BETWEEN POINT DEFECTS IN 2D**

An intriguing question is how point defects interact with each other in a 2D crystal. There have been no prior studies on this question to our knowledge. Here we attempt a preliminary investigation into this subject.

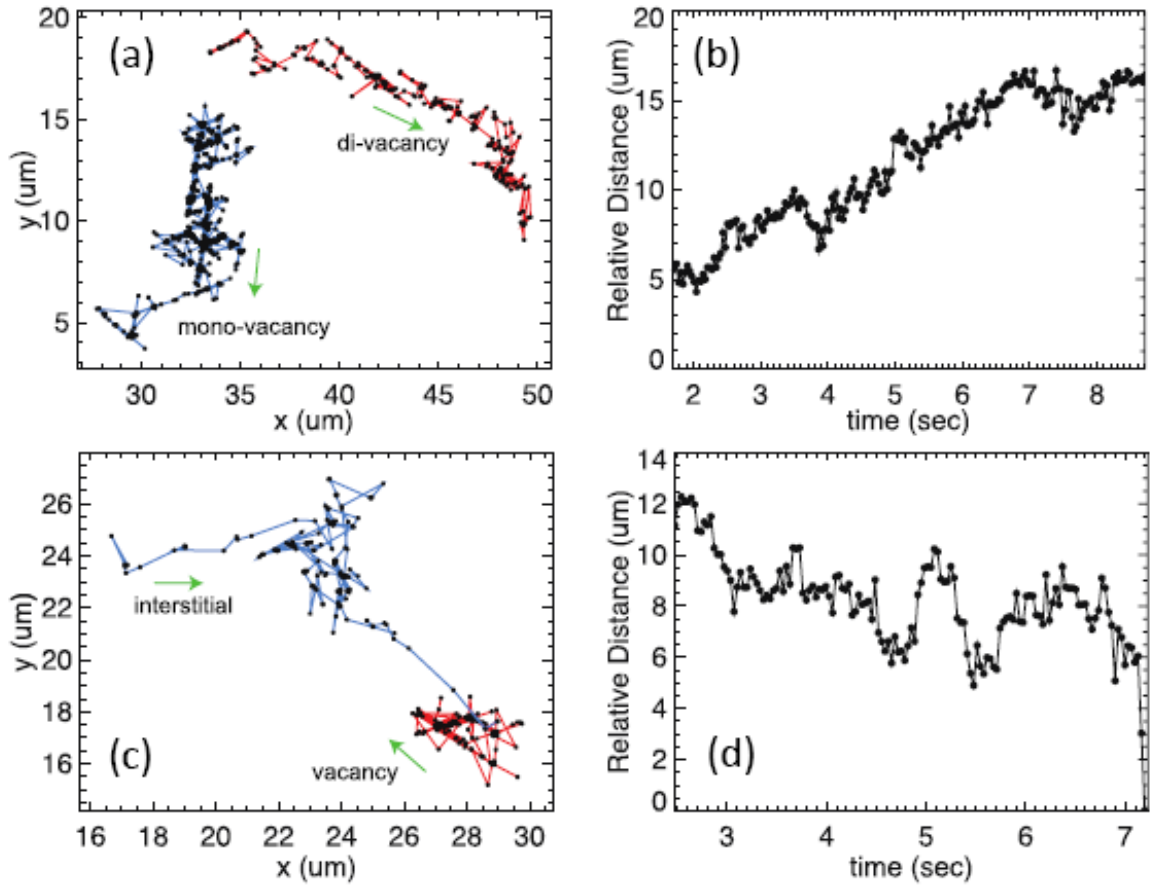

**Figure 3**: (Color online) (a) and (b) The trajectories of a mono-vacancy and a di-vacancy, and their relative distance vs. time; (c) and (d) The trajectories of a mono-vacancy and an interstitial, and their relative distance vs. time.

For the pair of mono-vacancy and di-vacancy, if any interaction existed between them, it appears to be repulsive, as their relative distance appears to grow almost linearly with time. An interesting feature of di-vacancy is that even though it occasionally breaks into a pair of dislocations, it never becomes two separate mono-vacancies.

These observations suggest that the vacancies are strongly attractive to each other when they are within a lattice constant in separation, but their interaction becomes repulsive once they are a few lattice constants apart.

As for the vacancy-interstitial pair, their relative distance is clearly decreasing with time until they crash into each other, and annihilate, as shown in Fig. 3(d), approximately 7 seconds after being created.

In Fig. 2, in (d) and (e), we saw that the interstitial moves by breaking into pairs of edge dislocations in a leap-frog fashion. This motion corresponds to the long jumps in Fig. 3 (c) for the interstitial. This observation suggests a simple mechanism for the interactions between the vacancy and interstitial. In thermal equilibrium, both vacancy and interstitial fluctuate between configurations that can be best described as pairs of edge dislocations with strain fields extending to large distances. Naively, one might expect the vacancy and interstitial to experience no interaction if separated by a few lattice constants, for both being point defects. However, the fluctuating strain fields may create an effective long-range attraction between the vacancy and the interstitial, since at distances far away their strain fields must cancel each other. It would be an interesting future experiment to determine the detailed forms of the effective potential between a vacancy and an interstitial, and between two vacancies.

## TRI-VACANCY: FORMATION AND DYNAMICS

Vacancies are suspected to play a central role in some models of melting [6]. For example, it has been proposed that a cluster of vacancies can form a nucleus of a droplet for nucleating a liquid phase. Of course, such models already assume a first-order phase transition scenario, which is still being debated.

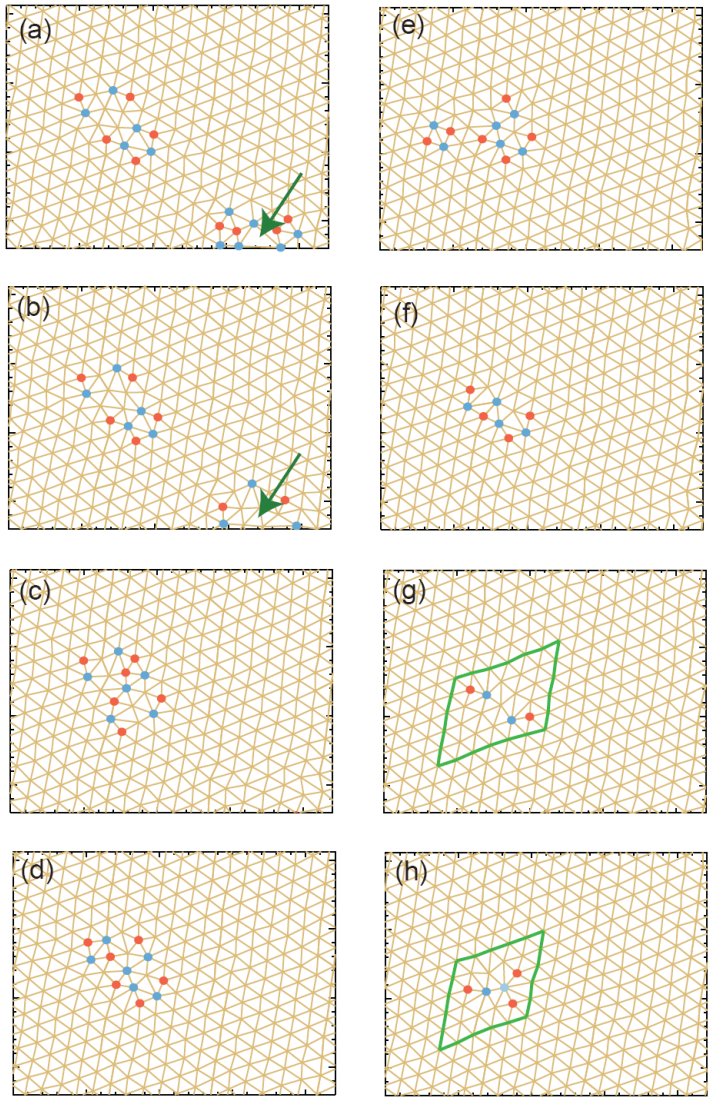

**Figure 4**: (Color online) The Delaunay triangulation diagram of the time-lapsed images of a tri-vacancy. The green arrows in (a) and (b) indicate the positions of the optical tweezers.

Here, we investigate the behavior of a tri-vacancy which can be viewed as a cluster of three mono-vacancies. As shown in Figure 4, we show the creation of a tri-vacancy by using the same technique. The green arrows in Fig. 4 (a) and (b) indicate the positions of the optical trap. The clusters of red and blue dots indicate the disclinations that make up a tri-vacancy.

After its creation, the local state around a tri-vacancy can be best viewed as a random assembly of particles, i.e. an amorphous or liquid-like state. After sometime, as shown in Fig. 4 (f) and (g), the defect structure resembles those of dislocation pairs.

But from time to time, the defect configuration fluctuates back into a state of random assembly, i.e., a liquid droplet, however, only in transient. We are in the process of looking into this phenomenon in more depth, with the goal of deciding the most probable configuration(s) of a tri-vacancy.

If the most probable configuration of a tri-vacancy is that of a liquid-droplet, a first-order melting scenario will be favored. On the other hand, if the most probable configuration is that of dislocation pairs, one may expect a continuous melting transition as in the theories of Kosterlitz and Thouless [7], and of Halperin and Nelson [8].

## SUMMARY AND DISCUSSIONS

We report additional studies of defect-creation using optical tweezers technique in 2D colloidal crystals. We present preliminary studies on the interactions between a vacancy-vacancy pair and a vacancy-interstitial pair. The results suggest that the vacancy-vacancy interaction appears to be repulsive (when they are a few lattice constants apart), while the interaction between a vacancy and an interstitial is attractive. A tri-vacancy is shown for the first time. It appears that the tri-vacancy fluctuates between an amorphous state and that of dislocation pairs. The relevance of this study to 2D melting is pointed out.

**Acknowledgement**: We benefitted from numerous discussions with Professor J.M. Kosterlitz. This work was supported by NSF-DMR under Grants No. 1005705 and No. 9804083.

*Present address: Memorial-Sloan Kettering Cancer Center, 411 East 67th Street, New York, NY 10065